\documentclass{ws-procs975x65}

\begin{document}

\title{A comparative study of the timing and the spectral properties during two recent outbursts 
(2010 \& 2011) of H 1743-322}

\author{DIPAK DEBNATH$^{1*}$, SANDIP. K. CHAKRABARTI$^{2,1}$, ANUJ NANDI$^{3,1}$}

\address{1. Indian Centre For Space Physics, 43 Chalantika, Garia Station Road, Kolkata, 700084, India\\
2. S.N. Bose National Center for Basic Sciences, JD-Block, Salt Lake, Kolkata, 700098, India\\
3. Space Astronomy Group, ISRO Satellite Centre, HAL Airport Road, Bangalore, 560017, India\\
}

%
%
%
%
%

\begin{abstract}

The Galactic black hole candidate (BHC) H~1743-322 recently exhibited two outbursts in X-rays in 
August 2010 \& April 2011. The nature (outburst profile, evolution of quasi-periodic oscillation (QPO) 
frequency and spectral states, etc.) of these two successive outbursts, which continued for around two 
months each, are very similar. We present the results obtained from a comparative study on the temporal 
and the spectral properties of the source during these two outbursts. The evolutions of QPOs observed in 
both the outbursts were well fitted with propagating oscillatory shock (POS) model. During both the 
outbursts, the observed spectral states (i.e, hard, hard-intermediate, soft-intermediate and soft) follow 
the `standard' type of hysteresis-loop, which could be explained with two component advective flow (TCAF)
model.


\end{abstract}


\bodymatter

\section{Introduction}

%

The Galactic transient BHC H~1743-322 showed several X-ray outbursts in the interval of $1-2$ years 
after its re-discovery in 2003 by Revnivtsev et al. (2003)\cite{Revnivtsev03}, 
after staying dormant for more than two decades. 
Recently in 2010 and 2011, H~1743-322 again found active in X-rays 
and was monitored with RXTE satellite on a daily basis. We make a comparative study of these two 
outbursts by studying temporal and spectral properties of the source using RXTE PCA archival data. 
The detailed results of this work are already in Debnath et al. (2013)\cite{DD13}.

\section{Observation and Data Analysis}

In 2010, $49$ observational IDs starting from 2010 August 9 (MJD = 55417) to 2010 September 30 
(MJD = 55469) are analyzed, 
and in 2011, $27$ observational IDs spread over the entire outburst, starting from 2011 April 12 to 2011 
May 19 (MJD = 55700) are analysed. 
We follow the standard data analysis techniques as presented in Nandi et al. (2012)\cite{Nandi12} 
using software package HEAsoft 6.12. 

\begin {figure}[t]
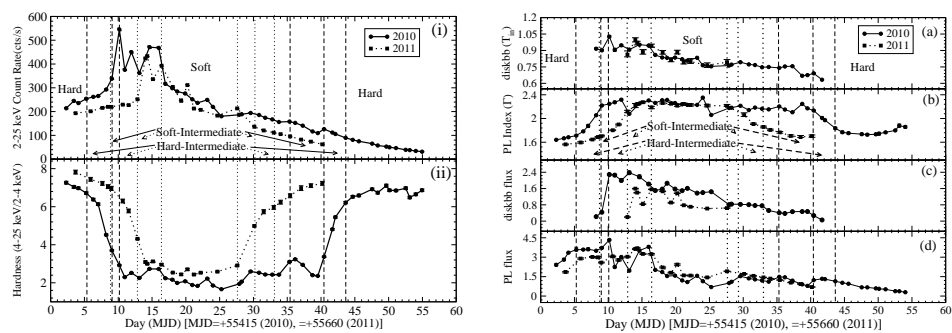

\vskip -0.5 cm
\centering{
\includegraphics[scale=0.6,angle=0,width=6truecm]{comb_pca-lc_2010-2011.eps}\hskip 0.5cm
\includegraphics[scale=0.6,angle=0,width=6truecm]{comp_flux_2010-2011.eps}}
\caption{(i) 2-25 keV PCA light curves and (ii) hardness ratios (4-25 keV versus 2-4 keV count ratio)
as a function of the MJD are shown in Left panel. In right panel, variation of the model fitted spectral 
components (diskbb and power-law) and their fluxes in 2-25 keV are shown. Vertical dashed and dotted lines 
indicate spectral state transitions for 2010 \& 2011 outbursts respectively.}
\label{kn : fig1}
\end {figure}

\section{Results - Temporal \& Spectral}
%


For studying temporal properties of H~1743-322 during 2010 \& 2011 outbursts, we extracted
light curves from PCU2 data of RXTE/PCA instrument in different energy bands: A: $2-4$ keV, 
B: $4-25$ keV, and C: $2-25$ keV. 
In left panel of Fig. 1, the variation of X-ray intensity in $C$ band and hardness ratio 
($B/A$ count rates) with MJD are shown. As of other transient BHCs (for e.g. GRO J1655-40, 
GX 339-4 etc.), during both the rising and the declining phases of these two outbursts, the 
evolution of low frequency QPOs are observed. The POS model fitted QPO frequency evolutions 
are discussed in Debnath et al. (2013)\cite{DD13}.


$2.5-25$ keV RXTE/PCA spectra are fitted with combination of thermal (diskbb) and non-thermal (power-law)
models, except data from initial and final hard and hard-intermediate spectral states (see right panel of 
Fig. 1). During the entire outbursts, 
evolutions of four observed basic spectral states in a same sequence: hard $\rightarrow$ hard-intermediate 
$\rightarrow$ soft-intermediate 
$\rightarrow$ soft $\rightarrow$ soft-intermediate $\rightarrow$ hard-intermediate $\rightarrow$ hard, 
are seen to form a hysteresis loop.

\section{Discussions and Concluding Remarks}

The evolution of QPO frequency can be explained by the movement of shock 
wave towards the BH
(rising phases) and away from the BH (declining phases). According to TCAF model\cite{CT95}, 
initially spectra are dominated by low-angular momentum sub-Keplerian flows and as a result, 
the spectra are hard. As the day progresses, rate of the thermally cool Keplerian matter increases and
the spectra become softer, progressively through hard-intermediate (Keplerian rate slightly less than
the sub-Keplerian rate), soft-intermediate (Keplerian rate comparable to the sub-Keplerian rate) and soft
state (dominating Keplerian rate). When viscosity is turned off at the outer edge, the declining phase
begins, and the Keplerian rate starts decreasing, and the spectra start to become harder again. 
However, the spectrum need not be retracing the same path, since the information about
the decrease of viscosity had to arrive at the viscous time scale. This causes a hysteresis effect. 




\begin{thebibliography}{9}
\bibitem{CT95} Chakrabarti, S.K. \& Titarchuk, L.G., {\em ApJ}, {\bf 455}, 623 (1995)
\bibitem{DD10} Debnath, D., Chakrabarti, S. K., \& Nandi, A., {\em A$\&$A} {\bf 520}, 98 (2010)
\bibitem{DD13} Debnath, D., et al., {\em ASR} (in press) (2013) (arXiv: astro-ph/1309.2564)
\bibitem{Nandi12} Nandi, A., Debnath, D., \& Mandal, S., et. al., {\em A$\&$A} {\bf 542}, 56 (2012)
\bibitem{Revnivtsev03} Revnivtsev, M., Chernyakova, M., \& Capitanio, F., et al., {\em ATel}, {\bf 132}, 1 (2003)

\end{thebibliography}

\end{document}